\documentclass[aps,prl,superscriptaddress,twocolumn,10pt]{revtex4-2}

\usepackage[titletoc,toc,title,page]{appendix}

\usepackage{csquotes}
\usepackage[italian,english]{babel}

\usepackage{microtype} 
\usepackage[margin=.6in]{geometry}

\usepackage[sc]{mathpazo} 

\usepackage{bm} 

\usepackage{dsfont} 
\usepackage{amsmath,amssymb,amsthm,thmtools}
\usepackage{mathtools}
\usepackage{cases}
\usepackage{calc}
\usepackage[usenames,dvipsnames]{color}
\usepackage{mathrsfs} 
\usepackage[normalem]{ulem} 
\usepackage{nameref}
\usepackage[colorlinks=true]{hyperref}
\usepackage[nameinlink]{cleveref}
\crefname{appsec}{Appendix}{Appendices}
\crefname{box}{Box}{Box}
\hypersetup{
  colorlinks   = true, 
  urlcolor     = green!80!black, 
  linkcolor    = blue, 
  citecolor    = red!80!black, 
  breaklinks=true   
}

\usepackage{physics} 

\usepackage{float} 

\usepackage{graphicx}

\usepackage[usenames,dvipsnames,table]{xcolor}
\usepackage{easyReview}

\usepackage{tikz}
\usetikzlibrary{calc,shapes.geometric}
\usepackage{amsmath,amssymb,amsthm,thmtools}
\usepackage{multirow,tabularx,booktabs}
\usepackage{makecell}
\usepackage[multiple]{footmisc}

\usepackage{lipsum} 

\makeatletter
 \preto\maketitle{
  \begingroup\lccode`~=`,
  \lowercase{\endgroup
  \let\saved@breqn@active@comma~
  \let~}\active@comma 
}
 \appto\maketitle{%
   \begingroup\lccode`~=`,
  \lowercase{\endgroup
   \let~}\saved@breqn@active@comma 
}
\makeatother

\newcommand{\parTitle}[1]{\noindent\emph{#1} --- }

\makeatletter
\newcommand{\labeltarget}[1]{\Hy@raisedlink{\hypertarget{#1}{}}}
\makeatother
\begin{document}

\title{A robust approach for time-bin encoded photonic quantum information protocols}

\author{Simon J. U. White}
\thanks{These authors contributed equally to this work.}

\author{Emanuele Polino\textcolor{blue}{$^*$}}

\email{e.polino@griffith.edu.au}

\author{Farzad Ghafari}
\email{f.ghafari@griffith.edu.au}
\author{Dominick J. Joch}
\author{Luis Villegas-Aguilar}

\affiliation{Centre for Quantum Dynamics and Centre for Quantum Computation and Communication Technology,  Griffith  University, Yuggera Country,   Brisbane,  Queensland,  4111  Australia}

\author{Lynden K. Shalm }
\author{Varun B. Verma}
\affiliation{National Institute of Standards and Technology,  325 Broadway,  Boulder, Colorado 80305, USA}

\author{Marcus Huber}
\affiliation{Atominstitut Technische Universität Wien, Stadionallee 2 1020, Vienna, Austria}
 \affiliation{Institute for Quantum Optics and Quantum Information (IQOQI), 
Austrian Academy of Sciences, Boltzmanngasse 3,  1090 Vienna,  Austria}

\author{Nora Tischler}

\email{n.tischler@griffith.edu.au}

\affiliation{Centre for Quantum Dynamics and Centre for Quantum Computation and Communication Technology,  Griffith  University, Yuggera Country,   Brisbane,  Queensland,  4111  Australia}

\begin{abstract}
Quantum states encoded in the time-bin degree of freedom of photons represent a fundamental resource for quantum information protocols. 
Traditional methods for generating and measuring time-bin encoded quantum states face severe challenges due to optical instabilities, complex setups, and timing resolution requirements. 
To circumvent these issues, we leverage an approach based on Hong-Ou-Mandel interference and we propose and demonstrate a robust and scalable protocol to generate and measure arbitrary high-dimensional time-bin quantum states. 
We experimentally implement the protocol in a photonic setup reaching high-fidelity quantum state tomographies of two and three-dimensional quantum states encoded in time-bins with short temporal separation. 
We also certify intrasystem polarization-time entanglement of single photons through a nonclassicality test. 
The demonstrated approach enables access to high-dimensional states and tasks that are practically inaccessible with standard schemes, thereby advancing fundamental quantum information science and opening applications in quantum communication.

\end{abstract}
\maketitle

Photonic quantum technologies rely on the capability to manipulate and measure quantum states of photons~\cite{slussarenko2019photonic,flamini2018photonic}.
Among the variety of photonic degrees of freedom that can be used to encode quantum states, one of the most noise-resilient and reliable is time. This degree of freedom was first proposed by Franson to test quantum nonlocality~\cite{franson1989bell}. Here, quantum states can be encoded using the arrival time of photons, i.e. time bins, which offer an infinite Hilbert space to encode information in high-dimensional states~\cite{de2002creating,humphreys2013linear,martin2017quantifying,erhard2020advances}. 
Besides their application to general quantum computation and simulation  tasks \cite{humphreys2013linear,takeda2019toward,dhand2018proposal,lubasch2018tensor}, time-bin states have been shown to be highly robust against turbulence and are suitable for long-distance transmissions in fibers~\cite{tittel1998violation,marcikic2003long,marcikic2004distribution,cuevas2013long,sun2016quantum}, free-space interconnects~\cite{schmitt2007experimental,ursin2007entanglement,fedrizzi2009high,jin2010experimental,steinlechner2017distribution}, and ground-satellite links~\cite{vallone2016interference,sidhu2021advances}.
 These properties make time-bin encoding a promising choice for quantum communication
\cite{brendel1999pulsed,tittel2000quantum,gisin2007quantum,zhong2015photon,islam2017provably,cozzolino2019high,fitzke2022scalable,kanitschar2024harnessing,valivarthi2020teleportation,wen2022realizing} and may be the foundation for the future quantum internet~\cite{wehner2018quantum,xu2020secure,diamanti2016practical}.

An increasingly important challenge is the robust and efficient generation and measurement of time-bin encoded states. To date, the most common techniques are based on unbalanced interferometers followed by time-resolved detection~\cite{franson1989bell,takesue2009implementation}, using post-selection for the recombination after an interferometer, or active switching protocols~\cite{vedovato2018postselection}. Unfortunately, interferometric techniques can face severe limitations, requiring large arm-length differences for time-bin resolution.
Although state-of-the-art superconducting single-photon detectors and electronics can reach a jitter and time resolution on the order of a few picoseconds \cite{esmaeil2017single,islam2017provably,korzh2020demonstration,taylor2022mid,hao2024compact}, standard technology and electronics available to most laboratories limit the overall time resolution to the order of nanoseconds. Consequently, the use of large unbalanced interferometers is often practically unavoidable. This leads to poor scalability in optical setups due to phase instabilities during long integration measurements, and impractical setups for high-dimensional states (requiring larger interferometers exacerbates these difficulties).
Other time-bin measurement techniques do exist, such as those based on nonlinear optical interactions, including pulse-shaping in conjunction with sum-frequency generation~\cite{pe2005temporal,kuzucu2008time,donohue2013coherent,maclean2018direct}. Alternatively, using both optical nonlinearity and interferometers allows one to resolve short time delays~\cite{bouchard2022quantum,bouchard2023measuring}. 
These approaches, however, come at the cost of increased complexity and expense, limiting their suitability for most applications.
To avoid the use of unbalanced interferometers, frequency-based techniques can be employed, but they only allow for particular measurements in the time-bin space, such as mutually unbiased bases~\cite{chang2023experimental}.
As a result, the practical drawbacks of standard measurement techniques pose severe limitations on the robustness, feasibility, and accessible Hilbert space dimensionality of time-bin states used in real experiments.

In this work, we employ a measurement scheme based on Hong-Ou-Mandel (HOM) interference~\cite{hong1987measurement} that overcomes the intrinsic limitations present in standard approaches and enables different tasks.
The HOM effect is based on the quantum interference between indistinguishable photons and is a building block for realizing interaction between photons in multi-photon protocols~\cite{bouchard2020two}, including those involving high-dimensional states \cite{zhang2021quantum}. 
Here, we apply the HOM measurement scheme to harness time-bin quantum encoding in a robust way
: That is, the scheme enables short temporal delays manipulated by reliable, versatile, and low-noise setups.

To generate arbitrary high-dimensional qudits in time, we propose a scheme based on quantum walk dynamics. When combined with the HOM measurement protocol, this approach allows for arbitrary measurements and circumvents traditional complexities associated with high-dimensional time-bin measurements. We perform optical experiments to demonstrate the reliability of both the state generation and measurement techniques.
First, we achieve high-quality quantum-state tomography, with an average fidelity between the reconstructed and target states of $\langle F \rangle = 0.9955 \pm 0.0007$ for qubits and $\langle F \rangle = 0.9912 \pm 0.0017$ for qutrits.
Next, we leverage the versatility of the protocol to perform a contextuality test for the intrasystem entanglement between the polarization and time degrees of freedom of single photons. 
We violate a Bell-like inequality and unambiguously demonstrate the presence of entanglement between these degrees of freedom. 
Finally, we discuss how these techniques can be extended to entangled systems with applications such as nonlocality tests or quantum communication.
By improving the robustness, versatility, and accessible dimensionality of time-bin encoded states, this work represents an advance toward scalable quantum technology protocols.

\begin{figure}[ht!]
\centering
\includegraphics[width=\columnwidth]{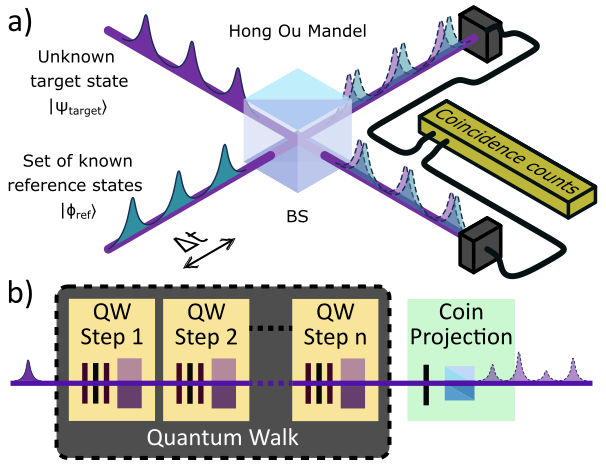}
 \caption{{\bf Conceptual scheme. a)} HOM-based measurement. An unknown target photon is measured by means of a known and controlled reference photon via HOM interference at a beam splitter (BS). From the coincidence counts, one can deduce the projection value.\textbf{ b)} State preparation via a linear discrete-time quantum walk (QW). Each QW step has polarization manipulation via waveplates and time-displacement via birefringent elements. A photon starting in a single time-bin can be probabilistically transformed to a time-bin qudit state.}
\label{fig:concept}
\end{figure}

\parTitle{Measurement scheme}
The goal here is to use HOM interference to perform arbitrary projective measurements on an unknown target time-bin encoded quantum state $\ket{\Psi_{\mathrm{target}}}$. 
Suppose the temporal wavefunction of a reference state is prepared in the state $\ket{\Phi_{\mathrm{ref}}}$. Then, one performs a HOM experiment (Fig.~\ref{fig:concept}a) interfering the two photons on a beam splitter (BS). The probability $P_{ab}$ of measuring a coincidence event between the two outputs of the BS (anti-bunching), given that the photons are indistinguishable in all other degrees of freedom besides time and path, will depend on their overlap~\cite{hong1987measurement,garcia2013swap}, which in the case of pure reference state reads:
\begin{equation}
P_{ab}= \frac{1- \left| \braket{\Phi_{\mathrm{ref}}} {\Psi_{\mathrm{target}}}\right|^2}{2} \; \;.
\label{eq:homoverlap}
\end{equation}

Arbitrary projections onto a target photon (either pure or mixed) can be realized by preparing a suitable (pure) reference photon and performing a HOM interference experiment. This type of measurement is general and can, in principle, be applied to any photonic degree of freedom (for polarization and hybrid polarization-time states see Refs.~\cite{tsujimoto2023quantum, Temporao2024, pilnyak2019quantum}, for frequency-temporal shape see \cite{wasilewski2007spectral,thiel2020single}). 
Additionally, various other types of light sources can serve as a reference, provided the overlap between the reference and target photonic wavefunctions (which may include non-time degrees of freedom) is properly accounted for in Eq.~\eqref{eq:homoverlap}~\cite{tsujimoto2023quantum}.
The minimum delay required to resolve two successive time-bins is fundamentally limited only by the coherence length of the single photons employed. This minimal delay maximizes the robustness against noise during preparation, transmission, and measurement.

The scheme can be directly generalized to arbitrary high-dimensional states because Eq.~\eqref{eq:homoverlap} provides information on the overlap between photonic states regardless of their dimensionality. Hence, assuming the ability to prepare arbitrary reference states, the measurement scheme applies to any time-bin encoded quantum state (qudit).
We also note that the error on overlap estimation through HOM interference is, in principle, independent of the dimensionality of the involved systems \cite{zhan2025experimental}.

\parTitle{Scheme for generating and measuring arbitrary high-dimensional time-bin photonic states}
In the context of state tomography, the HOM measurement scheme shifts the challenge of performing arbitrary projective measurements in high dimensions to the task of preparing high-dimensional reference states.
In order to do this, we propose an efficient probabilistic scheme to generate arbitrary high-dimensional quantum states via quantum walk (QW) dynamics~\cite{innocenti2017quantum}.
This approach has previously been realized in high-dimensional orbital angular momentum states~\cite{giordani2019experimental,suprano2021dynamical} and with 2D walks in the transverse momentum~\cite{esposito2023generation}.

The core tool for this task is a one-dimensional, discrete-time QW~\cite{aharonov1993quantum, venegas2012quantum} that involves two quantum systems: a walker implemented in the time-bin degree of freedom and a coin realized by the polarization. 
The QW dynamics (see Supplemental Material (SM)~II \cite{SI} ) can be used to prepare arbitrary qudit states in the walker space by choosing the optimal coin operations, corresponding to polarization operations, at each step~\cite{innocenti2017quantum}.
These dynamics produce a high-dimensional entangled target state in the coin-walker Hilbert space. The final pure walker state is probabilistically produced after projecting the coin state onto a suitable basis. Notably, it is always possible to optimize the set of polarization operations to maximize both the fidelity with an arbitrary target state and the success probability simultaneously~\cite{innocenti2017quantum, giordani2019experimental}. Furthermore, evidence suggests that the number of required steps grows only linearly with the dimension of the target state~\cite{innocenti2017quantum,giordani2019experimental}.

We consider the time degree of freedom of a photon as the walker, living in an infinite dimensional space $H_{\rm w}$ spanned by discrete ``position states'' i.e., time-bin states $\{ \ket{t_i}_\mathrm{w}\}$, with $i \in \mathbb{N}$.
The coin then corresponds to the polarization state of the photon, in a two-dimensional space $H_{\rm c}$ whose basis states, $\ket{\uparrow}\coloneqq\ket{H}$ and $\ket{\downarrow}\coloneqq\ket{V}$, denote the ``movement directions'' of the walker (see Refs.~\cite{schreiber2010photons,schreiber20122d,boutari2016large,lorz2019photonic,geraldi2021transient,fenwick2024photonic} for experimental realizations of time-bin-based quantum walks).

This polarization-time QW is implemented robustly using only waveplates and birefringent elements. Each step involves waveplates (quarter-half-quarter) setting the polarization (coin) state, followed by implementing a birefringent delay (temporal walk) exceeding the photons' coherence length. Finally, coin projection via a waveplate and PBS generates arbitrary time-bin states (see Fig.~\ref{fig:concept}b). Crucially, the QW can operate in a single spatial mode without spatial interferometers. The number of steps, $n$, determines the dimensionality ($n+1$), and adjusting waveplates alone modifies state generation or measurement, making this approach far more practical than spatial interferometric setups.

\begin{figure}[ht!]
\centering
\includegraphics[width=\columnwidth]{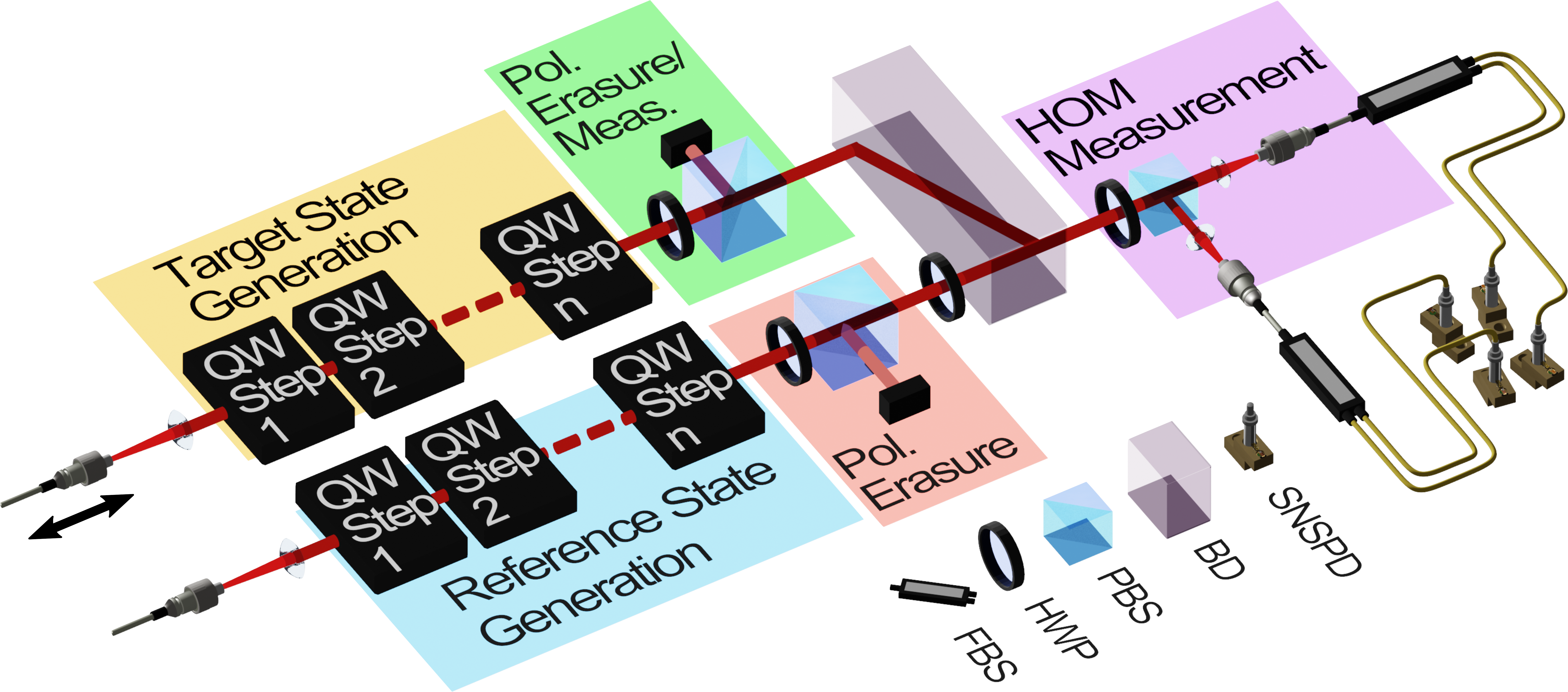}
 \caption{ {\bf Experimental setup.} The states of two single photons are independently encoded into time-polarization states using quantum walks (yellow and blue panels). The polarization 
 is either erased for the tomography experiment, using a HWP at $22.5^\circ$ and a PBS (red and green panels), or measured to verify intrasystem entanglement using a suitably rotated HWP and a PBS (green panel). Target and reference photons are recombined to perform projective measurements via HOM interference (purple panel); coincidences are recorded using SNSPDs and counting modules.   HWP, half-waveplate; PBS, polarizing beam splitter; BD, beam displacer; SNSPD, superconducting-nanowire single-photon detector; FBS, fiber-beam splitter.
 }
\label{fig:exp}
\end{figure}

\parTitle{Experimental demonstration for qubit and qutrit states} 
To verify the experimental performance of the scheme depicted in Fig.~\ref{fig:concept}, we carry out an experiment using single photons (see Fig.~\ref{fig:exp}). 
A pair of photons is generated using spontaneous parametric down-conversion (SPDC) by a pulse at time $t_0$, in a separable state of time and polarization $\ket{t_0}_s\ket{H}_s \otimes \ket{t_0}_i\ket{V}_i$, where the subscripts indicate signal and idler photon, respectively (see SM~III \cite{SI}). 
The signal and idler photons are split based on their polarization into paths 1 and 2, where the QW1 and QW2 are implemented to encode the target and reference states, respectively. To perform the HOM measurements, the photons are recombined in a single path and interfere via a final HWP at $22.5^\circ$ and PBS.

\begin{figure}[ht!]
\centering
\includegraphics[width=\columnwidth]{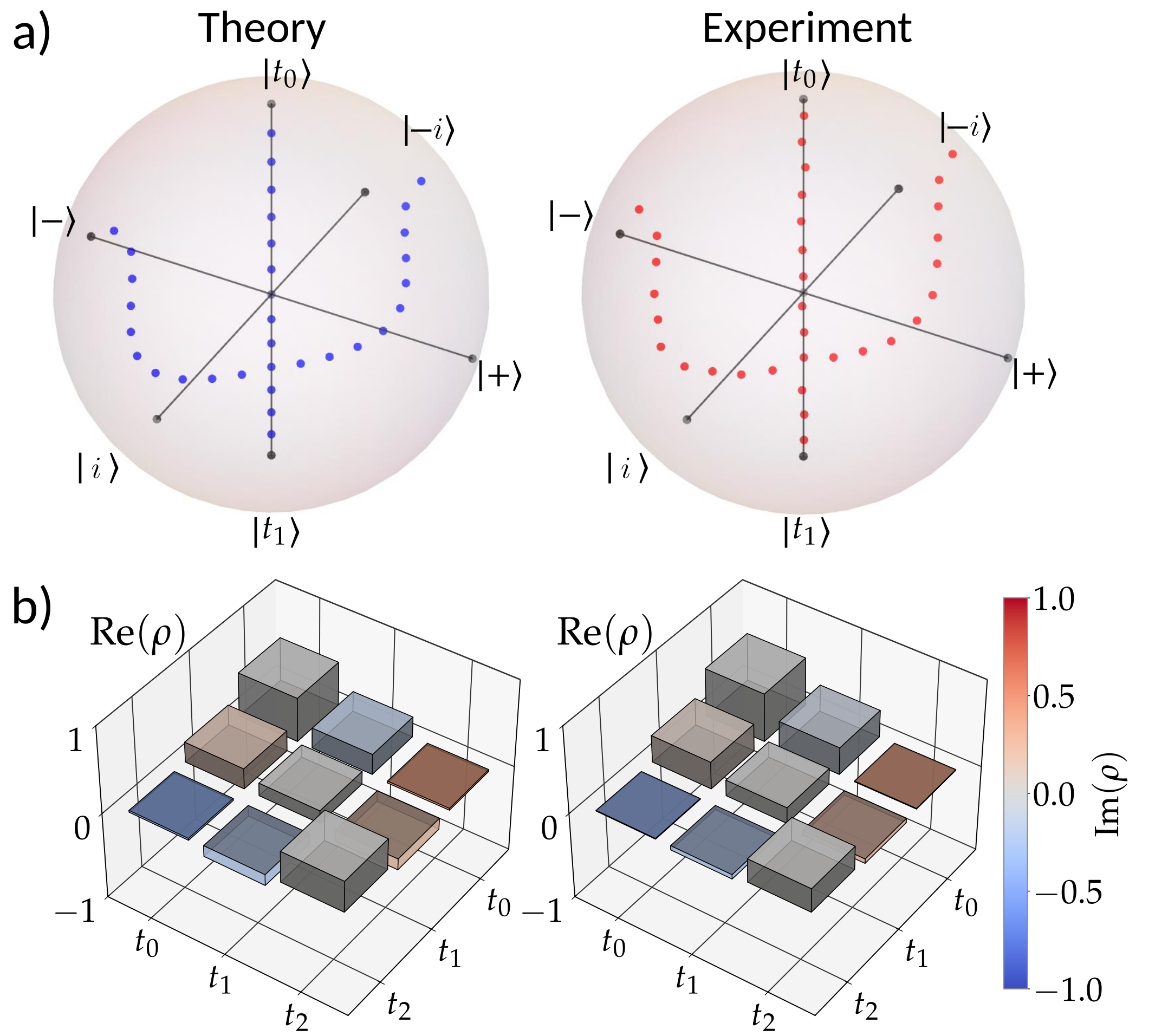}
 \caption{{\bf Experimental results.}
 {\bf a)}  Experimentally reconstructed mixed qubit states (red) and theoretical expectations (blue) on the Bloch sphere.
\textbf{b)} Example of the density matrix for a randomly prepared qutrit using the QW protocol---both the theoretical prediction and the experimental reconstruction.
The height of each bar and its label indicate the magnitude of the real part for each element of the density matrix.
The color of each bar corresponds to the magnitude of its imaginary component.}
\label{fig:results}
\end{figure}

To demonstrate the measurement scheme, we perform quantum state tomography of qubits and qutrits, implementing one- and two-step versions of the quantum walk experiments introduced above (see SM~II \cite{SI}).
For qubits, we use a single-step quantum walk (with a temporal delay of $\approx 8$ ps) and prepare 48 states, consisting of 15 approximately pure states and 33 states with varying degrees of mixture, as depicted in Fig.~\ref{fig:results}a. 
To generate the necessary mixture, we prepare pairs of pure states and periodically switch between them throughout the measurement.
For each tomography, we prepare the reference photon in the six states of the three mutually unbiased bases (MUBs). This measurement set is over-complete, but one can also use a smaller set employing different strategies (see SM~I \cite{SI}). We then reconstruct the target state using standard maximum likelihood estimation~\cite{james2001measurement}. For all pure and mixed states, we obtained a mean experimental fidelity~\cite{jozsa1994FidelityMixedQuantum} of $\langle F\rangle =0.9955\pm0.0007$ with respect to their theoretical targets (see SM for individual fidelities \cite{SI}). 

To showcase the generality of the scheme, we perform experimental quantum state tomographies for 3-dimensional quantum states, qutrits. Here, we realize two 2-step quantum walks with a temporal delay $\approx 5$~ps.  The reference states prepared are the $15$ linearly independent eigenvectors of the SU(3) generators, sufficient to reconstruct any unknown qutrit state~\cite{thew2002qudit}. We measured 18 approximately pure qutrit states and obtained an average fidelity of $\langle F\rangle =0.9912\pm0.0017$ with respect to the theoretical target states (see SM \cite{SI}). 
The error of the mean fidelity is estimated by calculating the standard deviation of the mean over the fidelities of all estimated states. This accounts for Poissonian and all the other fluctuations present in the generation and measurement of the states.
In Fig.~\ref{fig:results}{b} we present a reconstructed qutrit density matrix in a random superposition of the three basis states ($\ket{t_0}$, $\ket{t_1}$, and $\ket{t_2}$) and its theoretical target. The similarity in the amplitude of all components of the density matrix, including the color (representing the imaginary component), testifies to the high reliability and robustness of the generation and measurement scheme.

\parTitle{Contextuality test for hybrid entangled polarization-time single photon state}
The versatility of the scheme allows us to apply the measurement approach to quantum correlation tests.
Intrasystem entanglement describes the entanglement between different degrees of freedom of a single particle~\cite{shen2022nonseparable}.
These hybrid entangled states~\cite{simon2010nonquantum,eberly2016quantum,azzini2020single} hold promise for efficient high-dimensional quantum information schemes~\cite{forbes2019quantum}, as they can also be used to enlarge the overall Hilbert space of a single particle.
For photons, this kind of entanglement has been experimentally demonstrated through nonclassicality tests in orbital angular momentum-polarization states~\cite{karimi2010spin, karimi2014hardy,mclaren2015measuring, d2016entangled,milione2015using, cozzolino2019air, suprano2023orbital} and path-polarization~\cite{michler2000experiments, gadway2008bell}.

The experimental setup used for qubit tomography naturally produces time-polarization entangled single-photon states before the polarization erasure. We perform a nonclassicality test that unambiguously shows the presence of intrasystem entanglement in the time-polarization degrees of freedom through the violation of a Bell-like inequality. We carry out the time-bin measurements as described previously. To measure the polarization state of the single photons independently of the time, we adjust the polarization erasure step, i.e., the final polarization projection of the QW1 to instead perform projective measurements with the HWP and PBS (green panel Fig.~\ref{fig:exp}). For instance, if one prepares the photon in the $\ket{-}$ polarization, the single photon entangled state (before the polarization erasure) is:

\begin{equation}
\ket{\psi}=\frac{1}{\sqrt{2}}(\ket{H}\ket{t_0} - \ket{V}\ket{t_1}) \;.
\label{eq:intraent}
\end{equation}

In quantum mechanics, entanglement and nonlocality are distinct yet interconnected phenomena~\cite{werner1989quantum, brunner2005entanglement, luigi2024activation}. While the intrasystem entangled state in Eq.~\eqref{eq:intraent} cannot generate nonlocality due to the lack of spatial separation between measurements~\cite{khrennikov2020quantum}, it can give rise to a violation of classical noncontextual models~\cite{kochen1967problem, budroni2022kochen, paneru2020entanglement, karimi2010spin}. 

To demonstrate nonclassicality, and thereby certify time-polarization entanglement, we perform a Clauser Horne Shimony Holt (CHSH)~\cite{clauser1969proposed} type test, measuring the quantity $|S|=\left|\langle  A_0 B_0\rangle-\langle  A_0 B_{1}\rangle+\langle  A_{1} B_0\rangle+\langle  A_{1} B_{1}\rangle \right|$. Here, the correlation terms $\langle A_x B_y\rangle$ are calculated from the two-outcome measurements, $A_x$ and $B_y$, with binary measurement choices $x$ and $y$, performed on the time and polarization degrees of freedom, respectively. The relation $|S|\le 2$ must hold for any noncontextual classical model that describes the correlations between measurement outcomes.

After generating the state described in Eq.~\eqref{eq:intraent} and performing the optimal measurements on this state ($\{\sigma_z, \sigma_x \}$ for A, and $ \{ \frac{\sigma_x + \sigma_z}{\sqrt{2}}, \frac{\sigma_x - \sigma_z}{\sqrt{2}}\}$ for B), we obtain a value of $|S|^{\rm exp}=2.744 \pm 0.006$, where the error is estimated by considering Poissonian fluctuations of the counts, without rescaling the counts for accidental coincidences or limited HOM visibility.
This result indicates a violation of the noncontextual bound by over 100 standard deviations, unequivocally certifying the presence of intrasystem entanglement between the time and polarization degrees of freedom of the single photons.
Notably, the violation of the CHSH inequality also allows us to obtain a device-independent lower bound on the fidelity of the generated state (Eq.\eqref{eq:intraent}) via a self-testing procedure \cite{kaniewski2016analytic}. Using the analysis in \cite{kaniewski2016analytic}, we obtain a nontrivial bound on the fidelity $F \ge 0.941$ calculated for the worst-case scenario compatible with the experimental error.

\parTitle{Discussions}

Time-encoded quantum states are powerful tools for photonic quantum information. Standard techniques to manipulate and measure these states suffer from instabilities and detection limitations, restricting their practicality. 

In this work, we used a scheme based on HOM interference to overcome some of these limitations and perform proof-of-principle experiments, with high versatility and fidelity.
The scheme enables the use of time-bin qudits with minimal temporal delay between bins, limited only by the coherence time of the photons. Using femtosecond sources, this method could surpass existing temporal resolution limits. We experimentally generated and measured high-dimensional time-bin states (up to dimension 3) and detected nonclassical polarization-time entanglement of single photons.

As a perspective, the demonstrated scheme can be used for different tasks. One possible scenario is the preparation and measurement of a single photon in a high-dimensional state for general quantum communication.
Here, the sender uses one QW to prepare the message state to be sent, while the receiver uses a second QW to prepare the state of a reference photon, which is used to measure the message. The reference photon could be the twin photon generated by an SPDC source or a photon generated by a source synchronized with the source generating the first photon. The rapid development of on-demand single-photon sources, such as quantum dots, has already enabled several time-bin-encoded demonstrations~\cite{jayakumar2014time,takemoto2015quantum,yu2015two,lee2019quantum,anderson2020gigahertz,carosini2024programmable}, and such sources represent promising candidates for implementing high-dimensional single-photon time-bin states within our approach.

\begin{figure}[ht!]
\centering
\includegraphics[width=\columnwidth]{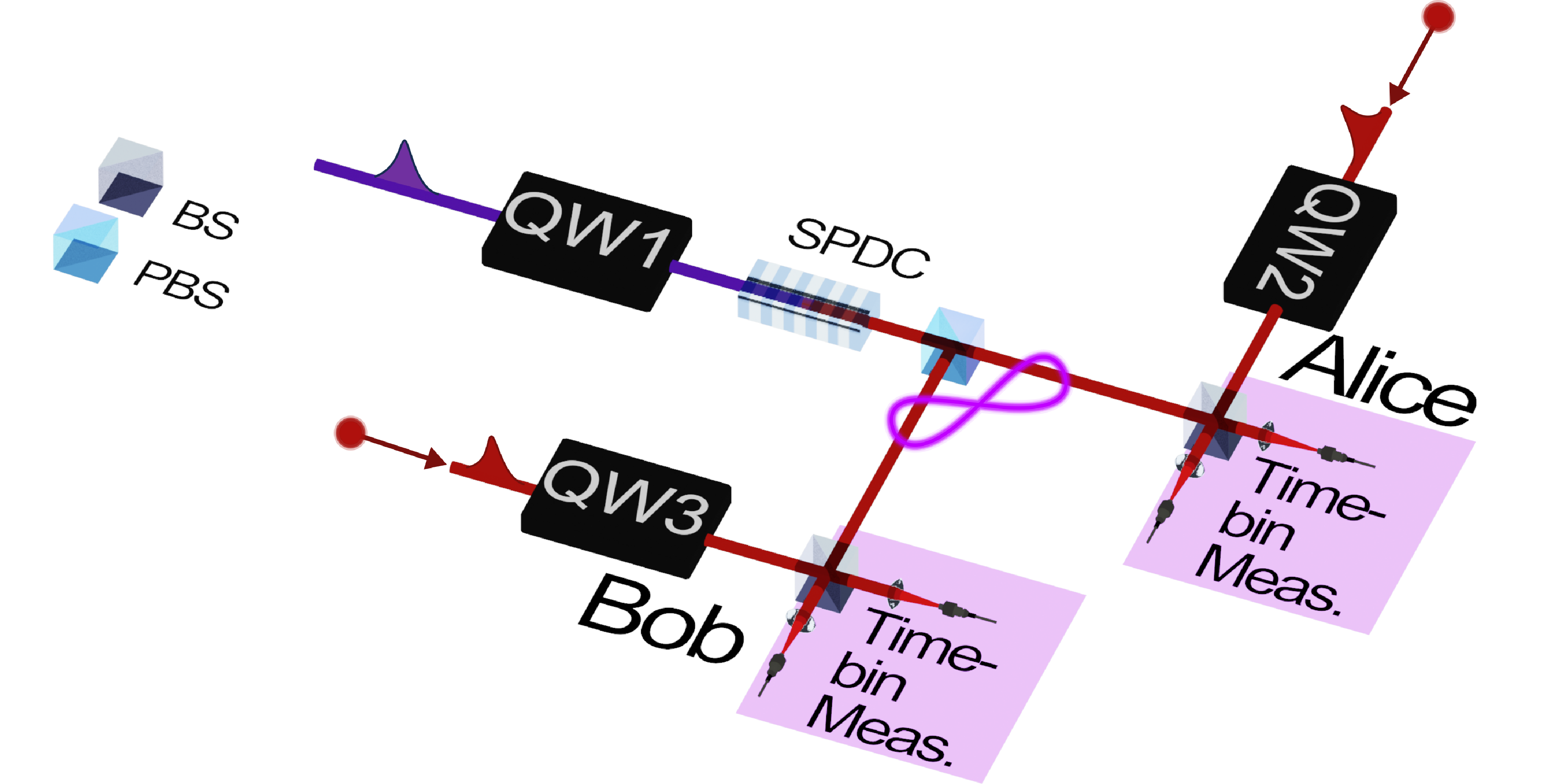}
\caption{
\textbf{Conceptual scheme for high-dimensional time-bin-based quantum key distribution}
Using the QWs as a building block, high-dimensional entangled states can be prepared, distributed, and measured. PBS, polarizing beam splitter; BS, beam splitter.} 
\label{fig:schemes}
\end{figure}

A second scenario would be high-dimensional quantum key distribution (QKD), depicted in Fig.~\ref{fig:schemes}, where a pair of time-bin entangled photons are shared among two measurement stations. The entangled state of the photons generated using an SPDC source can be engineered by suitably manipulating the pump in the time-bin degree of freedom. Here, a QW could be used to generate the time-bin state $\ket{\Psi_p}=\sum_{j=1}^n \; a_j \ket{t_j}$, where $a_j$ are arbitrary coefficients and $\sum_{j=1}^n |a_j|^2=1$. Considering the cases where the probability of generating a pair of photons in a pulse is very low (neglecting higher-order SPDC terms), the following entangled state will be generated: $\ket{\Psi}=\sum_{j=1}^n a_j \ket{t_j}_s\ket{t_j}_i $. The entangled photons are distributed to measurement stations and measured using reference photons from respective QWs (see Fig.~\ref{fig:schemes}, and SM Fig.~S3 \cite{SI}).

An attractive feature of these schemes is their scalability, as the number of optical elements scales linearly with the number of steps~\cite{innocenti2017quantum,giordani2019experimental}. Therefore, the effectively certifiable dimension of the state is fundamentally limited only by the ratio between the pump pulse separation and the coherence length of the generated photons and not by the time resolution of the detector.

 The generation and measurement of high-dimensional entangled states offer the potential for higher key rates \cite{kanitschar2024practical} and also address some of the fundamental challenges in quantum communication, such as enhancing the robustness of entanglement against noise~\cite{ecker2019overcoming,doda2021quantum,bulla2023nonlocal} 
 and enabling new methods for testing the foundations of quantum mechanics~\cite{vertesi2010closing,collins2002bell,vaziri2002experimental,dada2011experimental}. 
 If the time bins can be temporally resolved by the detection apparatus, a significant benefit for QKD is the ability to perform multi-outcome measurements (those encoding the exchanged key), yielding $\log(d)$ bits for each coincident event. The test measurements can be implemented with the help of QW and would only require a small fraction of total rounds without the usual restriction of unbalanced interferometers.
We note that the HOM-based measurement approach (though presented using photonic information carriers) can, in principle, be applicable to other physical platforms, including electrons, atoms, collective atomic excitations, plasmons, and phonons where HOM interference has been demonstrated~\cite{bouchard2020two}. 
In summary, by enhancing the robustness and enlarging the accessible dimensionality, this work provides a step towards the scalable implementation of high-dimensional time-bin qudits as quantum information carriers.

\begin{acknowledgments}
\textit{Acknowledgments--}
This work was supported by ARC Grant No.~CE170100012.
L.V.-A. and D.J.J. acknowledge support from the Australian Government Research Training Program (RTP).
We thank Sae-Woo Nam for supporting this work through his development of superconducting-nanowire single-photon detectors and Alex Pepper for helpful discussions and support with the experimental setup. M.H. acknowledges support from the Horizon-Europe research and innovation programme under grant agreement No.~101070168 (HyperSpace). 
\end{acknowledgments}


%

\clearpage
\onecolumngrid

\appendix
\makeatletter
\renewcommand{\thesection}{S\arabic{section}}
\renewcommand{\theequation}{S\arabic{equation}}
\renewcommand{\thefigure}{S\arabic{figure}}
\renewcommand{\thetable}{S\arabic{table}}

\def\@seccntformat#1{\csname the#1\endcsname\quad}
\makeatother

\setcounter{section}{0}
\setcounter{equation}{0}
\setcounter{figure}{0}
\setcounter{table}{0}

\thispagestyle{empty}  

\begin{center}
    \textbf{\large Supplemental Material for: Robust Approach for Time-Bin-Encoded Photonic Quantum Information Protocols}
\end{center}

\begin{center}
Simon J. U. White\textcolor{blue}{$^*$},$^{1}$
Emanuele Polino\textcolor{blue}{$^*$},$^1$
Farzad Ghafari,$^{1}$
Dominick J. Joch,$^{1}$
Luis Villegas-Aguilar,$^{1}$
Lynden K. Shalm,$^{2}$
Varun B. Verma,$^{2}$ 
Marcus Huber,$^{3,4}$
and  Nora Tischler$^1$
\end{center}
\vspace{-0.5cm}
\begin{center}
    \text{\small \it $^{1}$ Centre for Quantum Dynamics and Centre for Quantum Computation and Communication Technology,} \\
    \text{\small \it Griffith University,  Yuggera Country,   Brisbane, Queensland 4111, Australia}
    
    \text{\small \it $^{2}$ National Institute of Standards and Technology,  325 Broadway,  Boulder,  Colorado 80305, USA}

    \text{\small \it $^{3}$ Atominstitut, Technische Universität Wien, Stadionallee 2, 1020 Vienna, Austria}
 
    \text{\small \it $^{4}$ Institute for Quantum Optics and Quantum Information (IQOQI), Austrian Academy of Sciences, Boltzmanngasse 3, 1090 Vienna, Austria}
\end{center}
\vspace{-0.5cm}
\section{I. Different possible HOM measurement normalization techniques to estimate the antibunching probabilities}
\begin{figure*}[ht!]
\centering
\includegraphics[width=0.7\textwidth]{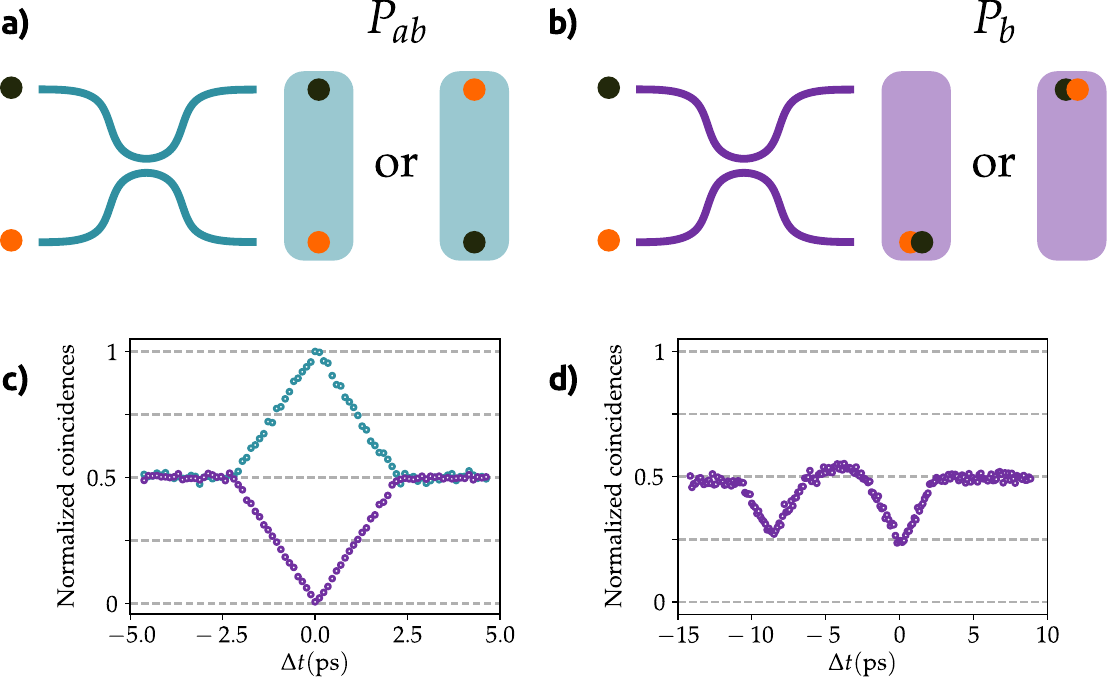}
\caption{{\bf HOM measurement events.} {\bf a)} Antibunching events. {\bf b)} Bunching events. {\bf c)} Experimental HOM scan where both bunching (purple points) and antibunching (cyan points) are measured. The input photons are indistinguishable and are both in the state $\ket{t_0}$. The antibunching statistics, as a function of the relative delay between the two photons, is estimated using fiber beam splitters connected to the outputs of the HOM beam splitter.  {\bf d)} Antibunching events of the HOM scan with input photons with states $\ket{t_0}$ and $\ket{+}=(\ket{t_0}+\ket{t_1})/\sqrt{2}$, respectively. Error bars are due to Poissonian statistics.
}
\label{fig:bunching}
\end{figure*}

To perform a quantum state tomography using HOM measurements, one must estimate the antibunching probability $P_{ab}$ in Eq.~(1) of the main text. Normalization from raw coincidence counts to retrieve the probability $P_{ab}$ can be performed in a number of ways, given that the setup described in the main text allows one to obtain both antibunching and bunching events in the experiment directly. 
Antibunching events refer to events where the two photons exit the beam splitter along two different output ports, while the bunching events are those where the photons exit the same output port (Fig.~\ref{fig:bunching} a) and b), respectively).

 In Fig.~\ref{fig:bunching} c), we show a measured HOM dip and peak, corresponding to bunching and antibunching respectively, resulting from interference of two photons in the same time-bin state $\ket{t_0}$. The HOM visibility obtained for indistinguishable photons in our setup amounts to $V=98.5 \pm 0.3\%$.
In Fig.~\ref{fig:bunching} d) we present an example of the data collected when interfering a time-bin superposition state $\ket{+}=(\ket{t_0}\pm\ket{t_1})/\sqrt{2}$, with the state $\ket{t_0}$. Here, we observe two HOM dips with reduced visibility $V\simeq0.5$, corresponding to the photon wavefunction split among two time bins.

\begin{figure*}[t!]
\centering
\includegraphics[width=1\textwidth]{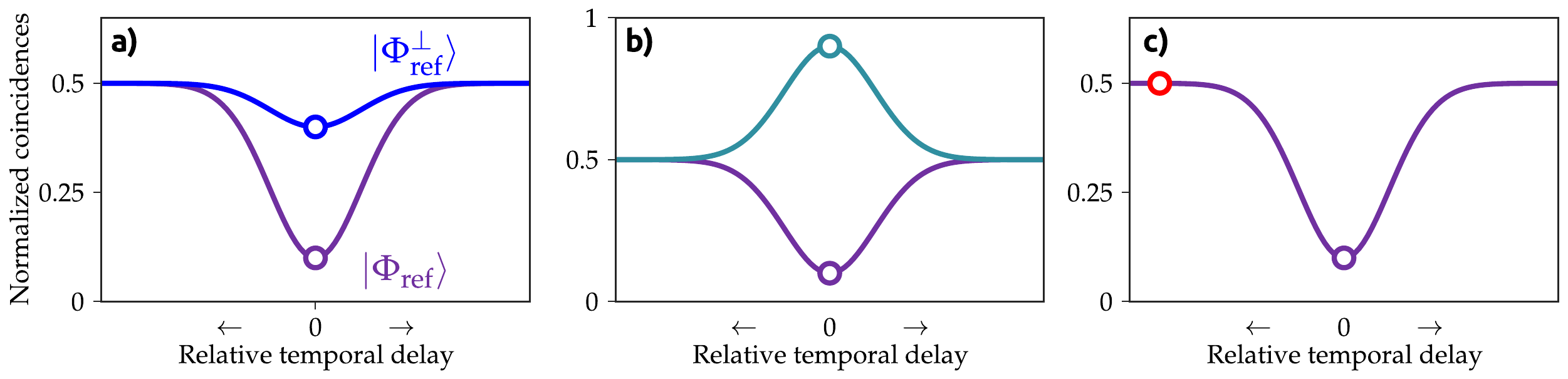}
\caption{{\bf Conceptual normalization methods.} Three methods for normalizing the antibunching coincidences of a HOM projection. The circles indicate the two measurements required to normalize the raw coincidences and obtain $P_{ab}$.
{\bf a)} Normalized antibunching counts ($P_{ab}$) corresponding to a projection with the reference state $\ket{\Phi_{\rm ref}}$ (bottom, purple) and its orthogonal complement $\ket{\Phi^{\perp}_{\rm ref}}$ (top, blue). Note that, inverting Eq.(1) of the main text, the overlap between target and reference photons is given by: $\left| \langle \Phi_{\rm{ref}}| \Psi_{\rm{target}} \rangle \right|^2 =1-2 P_{ab}$. Hence, the two probabilities $P_{ab}$ relative to the references $\ket{\Phi_{\rm ref}}$ and $\ket{\Phi^{\perp}_{\rm ref}}$, sum up to $0.5$ in the zero-delay point.
{\bf b)} Both the bunching (top circle) and antibunching (bottom circle) counts of the same HOM projection.
{\bf c)} Antibunching counts of a projection at the center of the HOM dip (central circle) and far from the interference region (left circle). When the relative temporal delay is large enough, the two photons are always distinguishable, regardless of their time-bin state.
}
\label{fig:measchemes}
\end{figure*}

In our experiment, we focused on estimating the probabilities using only antibunching coincidence counts. Specifically, for each measurement, we use the antibunching coincidences from the reference state and its orthogonal state(s). For example, in the qubit Hilbert space for the time computational basis one would measure the coincidences with the target and both $\ket{t_0}$ and $\ket{t_1}$ for normalization, as illustrated in Fig.~\ref{fig:measchemes} a). 
This allows us to project onto all six (for dimension two) mutually unbiased bases (MUBs), two for each basis, for the tomographic reconstruction of each state.

We briefly outline two alternative methods for normalizing the counts and retrieving the probability.
One approach, depicted in Fig.~\ref{fig:measchemes} b), involves normalizing the antibunching terms by estimating the bunching terms through photon-number-resolving detection.
Photon-number resolution can be achieved deterministically with specialized detectors, such as transition edge sensors, or probabilistically using standard detectors following beam splitters (pseudo-photon-number resolution).
This approach allows for the simultaneous measurement of both bunching and antibunching events, enabling single-shot measurements in a manner similar to standard dichotomic polarization measurements or high-dimensional holographic OAM measurements (see Refs.~[60,109,110] in the main text).
Such a capability opens new possibilities for applications, including quantum key distribution (QKD) protocols.
Nonetheless, accurately estimating the bunching probability requires precise characterization of the beam-splitter splitting ratios and detector efficiencies.
Another (non single-shot) technique to estimate the antibunching probability involves measuring the antibunching counts far outside the center of the HOM dip, where the photons are fully distinguishable regardless of their time-bin state (Fig.~\ref{fig:measchemes} c). 

In our experiment, we chose to estimate the overlaps using only antibunching coincidences, thus avoiding the need for beam splitter characterization and the need to move mechanical translation stages. The additional two fiber-based beam splitters at the outputs of the setup were exclusively used to monitor power fluctuations.
We highlight that the time requirements for this method are comparable to those of standard holographic techniques employed to measure OAM states of light \cite{dada2011experimental,giordani2019experimental,suprano2021dynamical}. This again represents an advantage over conventional interferometric techniques for measuring time-bin quantum states, particularly in high-dimensional scenarios, where altering the measurement basis would typically require substantial adjustments to large spatial interferometers.

\section{II. Discrete-time Quantum walk}

The protocol proposed in the main text to generate arbitrary high-dimensional time-bin states is based on the dynamics of discrete-time quantum walks (QWs).
A discrete-time quantum walk comprises two systems:  a walker and a coin. The general quantum state of the walker will be  $\ket{\rm walker}=\sum_i \, c_i\, \ket{t_i}_{\rm w}$, with $\sum_i |c_i|^2=1$, and the coin lives in a two-dimensional Hilbert space  $\mathcal{H}_{\rm c}$ with basis $\{\ket{\uparrow}\coloneqq\ket{H},\ket{\downarrow}\coloneqq\ket{V}\}$. In our proposal, the coin is the polarization, and the walker lives in the time-bin space.

The complete dynamics of discrete-time QW is determined by a global evolution unitary operator $U_t$ acting on the global system $\ket{\Psi}_{\rm tot}$ at each step $t$.
This unitary can be decomposed, at each step $t$, into a coin operator $C_t$ acting solely on the polarization (coin) state, followed by a shift operator $S_{\rm cw}$ that controls the walker's movement depending on the coin state \cite{venegas2012quantum}:
\begin{equation}
S_{\rm cw}=\sum_k |k\rangle \langle k|_{\rm w}\otimes |{\downarrow}\rangle \langle {\downarrow}|_{\rm c}+ |k+1\rangle \langle k|_{\rm w}\otimes |{\uparrow}\rangle\langle{\uparrow}|_{\rm c}\;,
\label{shift}
\end{equation}
where $k$ represents the position occupied by the walker in the time-bin space.
The shift operator $S_{\rm cw}$, which controls the movement of the walker depending on the coin state \cite{venegas2012quantum}, can be realized using a birefringent plate, which delays the time position of the photon according to its polarization. 
The global evolution at step $t$ is given by: $U_t=S_{\rm cw}\,(C_t\otimes \mathbb{I}_{\rm w})$, where $\mathbb{I}_{\rm w}$ is the identity in walker space.
Thus, the final state after $n$ steps will be
\begin{equation}
\ket{\Psi}^n_{\rm tot}=\prod_{t=1}^n S_{\rm cw}\,(C_t \otimes \mathbb{I}_{\rm w})\; \ket{\Psi}^0_{\rm tot}\;,
\label{eq:finalstateqw}
\end{equation}
where $\ket{\Psi}^0_{\rm tot}$ is the initial state of the global system.

In terms of loss, our protocol is scalable relative to standard schemes requiring an exponential increase of optical elements with the target dimension. There are two reasons for this in our scheme.\\
First, the number of optical elements required to generate a state scales linearly with the dimension of the target. This means that loss scales exponentially with dimension $d$ with an exponent that is only linear in $d$. \\ For example, if we need to generate a state with dimension $d$, the number of required elements will be $O(d)$, and assuming a fixed transmission  $\eta$ for each step of the QW, the overall transmission will scale as $\eta^{O(d)}$. \\
Second, the success probability of generating arbitrary reference states using the quantum walk scheme remains approximately constant, within numerical precision, as the dimension increases~\cite{innocenti2017quantum,giordani2019experimental}.

\section{III. Experimental details}

The single-photon source is realized using an 80 MHz pulsed laser centered at $775$~nm, pumping a 1.5~cm-long periodically poled potassium titanyl phosphate (PPKTP) crystal. Type-II spontaneous parametric down-conversion (SPDC) produces spectrally degenerate photon pairs at $1550$~nm. The photons are split based on their polarization and directed to the qubit/qutrit experiments via an optical fiber.

\subsection{A. Qubit generation and measurement setup}
\begin{figure}[ht!]
\centering
\includegraphics[width=0.7\columnwidth]{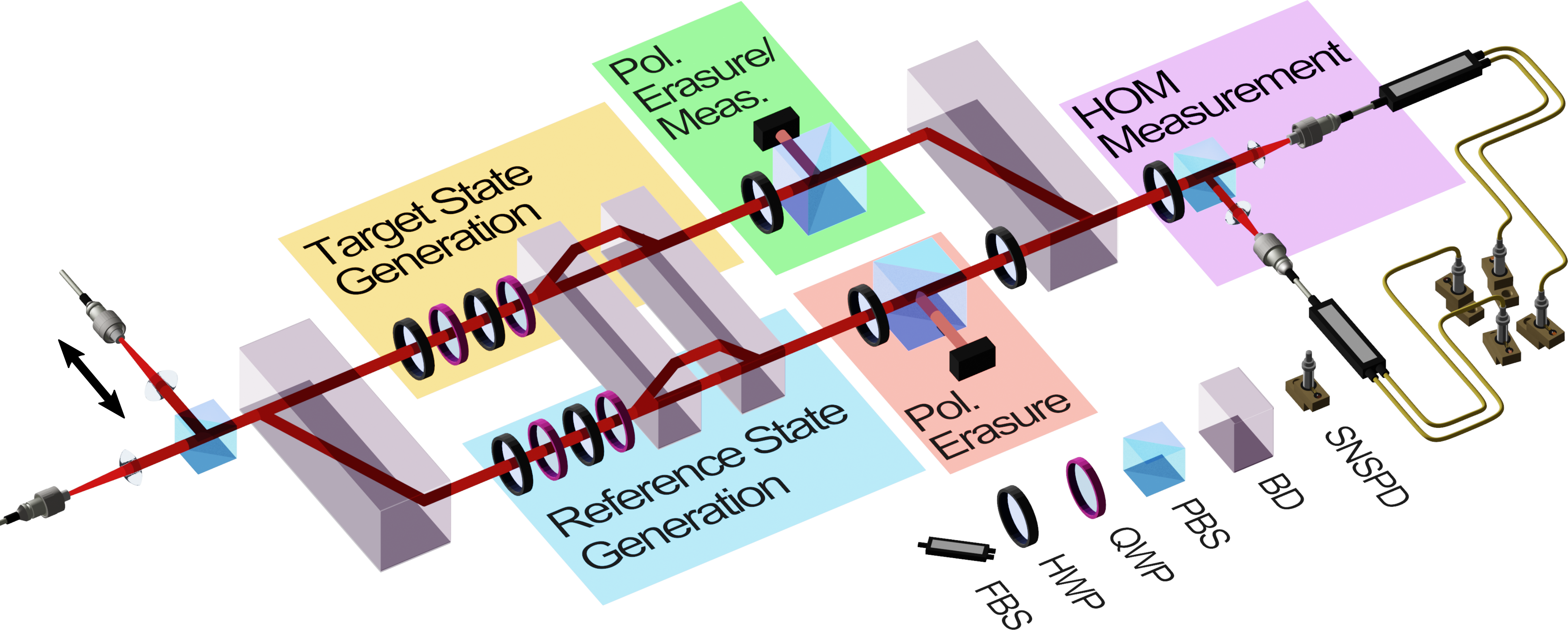}
 \caption{ {\bf Experimental setup for time-bin qubits.} The states of two single photons are independently encoded using their polarization, and are then mapped into time-polarization states (yellow and blue panels). The polarization 
 is either erased for the tomography experiment, using a PBS (red and green panels), or measured to verify intrasystem entanglement using a suitably rotated HWP and a PBS (green panel). Target and reference photons are recombined to perform projection via HOM interference (purple panel); coincidences are recorded using SNSPDs and counting modules.   HWP, half-wave plate; QWP, quarter-wave plate; PBS, polarizing beam splitter; BD, beam displacer; SNSPD, superconducting-nanowire single-photon detector; FBS, fiber beam splitter.
 } %
\label{fig:expqubits}
\end{figure}

For generating time-bin qubits, the relative temporal delay between $\ket{t_0}$ and $\ket{t_1}$ introduced by the interferometer is around $8$~ps, only slightly longer than the coherence time of the photons of $\approx 2.3$~ps (generated using a picosecond laser). Two sets of wave plates control the polarization of each photon independently. 
The temporal delays define a two-dimensional Hilbert space spanned by the two time-bin states $\{\ket{t_0},\ket{t_1} \}$.
A final beam displacer recombines the two time-encoded photons into the same spatial mode, after which a polarization-based HOM experiment (equivalent to the one in Fig.~1 in the main text)) is performed with a fixed HWP at $22.5^\circ$ and a PBS. Coincidences are then measured using superconducting nanowire single-photon detectors (SNSPDs) as shown in Fig.~\ref{fig:expqubits}.

\subsection{B. Qutrit generation and measurement setup}
The experimental configuration for producing qutrits resembles that used for qubits, except that state manipulation requires two two-step quantum walks (see Fig.~2 of the main text).
The time-bin shift operation of each quantum walk step is implemented using non-displacing PPKTP crystals. A key advantage of employing a birefringent crystal lies in its ability to perform the time-bin shift operation within a single spatial mode, thereby eliminating the need for unbalanced interferometers. The PPKTP crystals introduce a delay of approximately $4.5$ ns between the two polarizations of the single photons. For the qutrit demonstration, we utilize the shorter pump pulses of a femtosecond laser to generate photons with coherence lengths of $\approx0.8$~ps. 

\section{IV. Results of quantum state tomographies}

\begin{figure*}[ht!]
\centering
\includegraphics[width=0.8\textwidth]{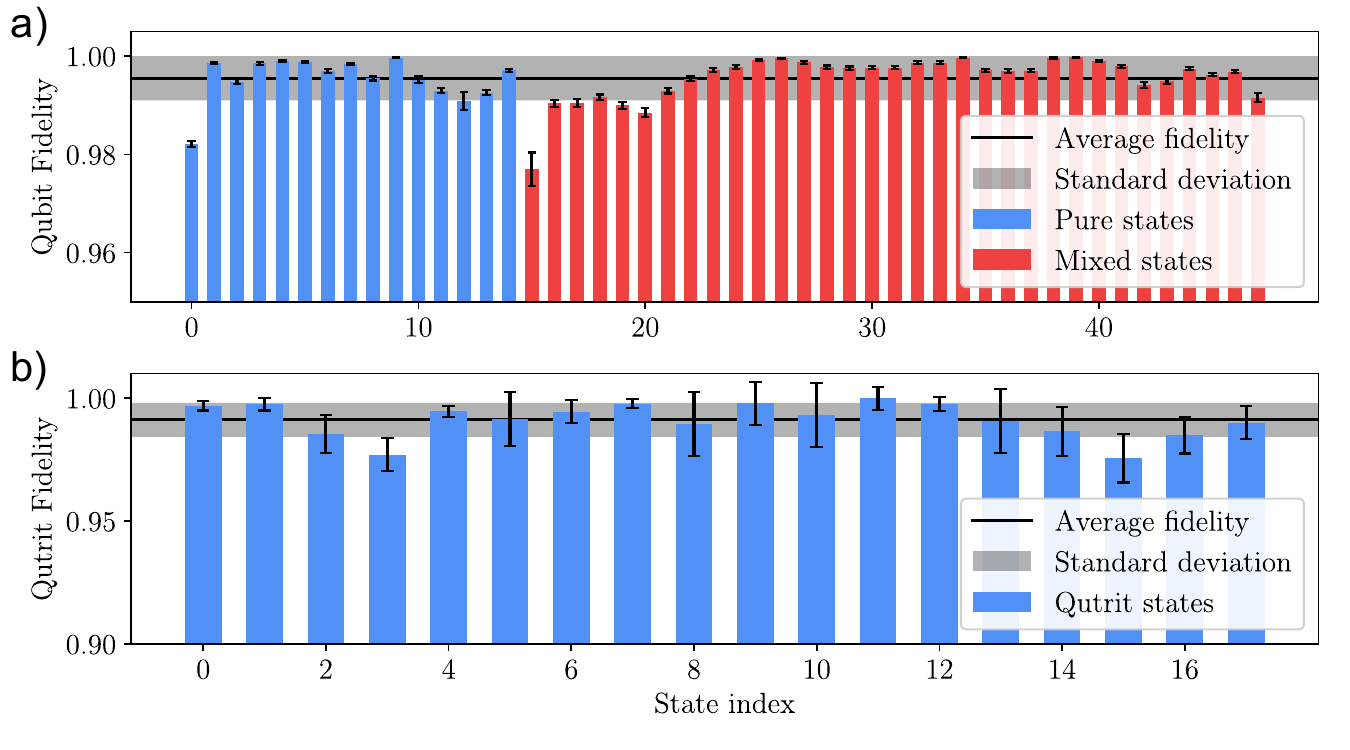}
\caption{{\bf Experimental fidelities of individually reconstructed states. }\textbf{a)} Fidelity values for 15 pure and 33 mixed qubit states. \textbf{b)} Fidelity values for 18 qutrit states. The 15 states correspond to the eigenvectors of the Gell-Mann matrices, which form a complete basis for the qutrit Hilbert space. The remaining three qutrit states are random superpositions of all three basis states ($\ket{t_0}$, $\ket{t_1}$, and $\ket{t_2}$).\\
The error bars of each fidelity represent  
the errors calculated from the experimental counts, assuming Poissonian statistics. The gray band instead is $\pm 1$ standard deviation calculated using all the fidelity values for the estimated states.
}
\label{fig:Fidelities}
\end{figure*}

\begin{figure*}[h!]
\centering
\includegraphics[width=\textwidth]{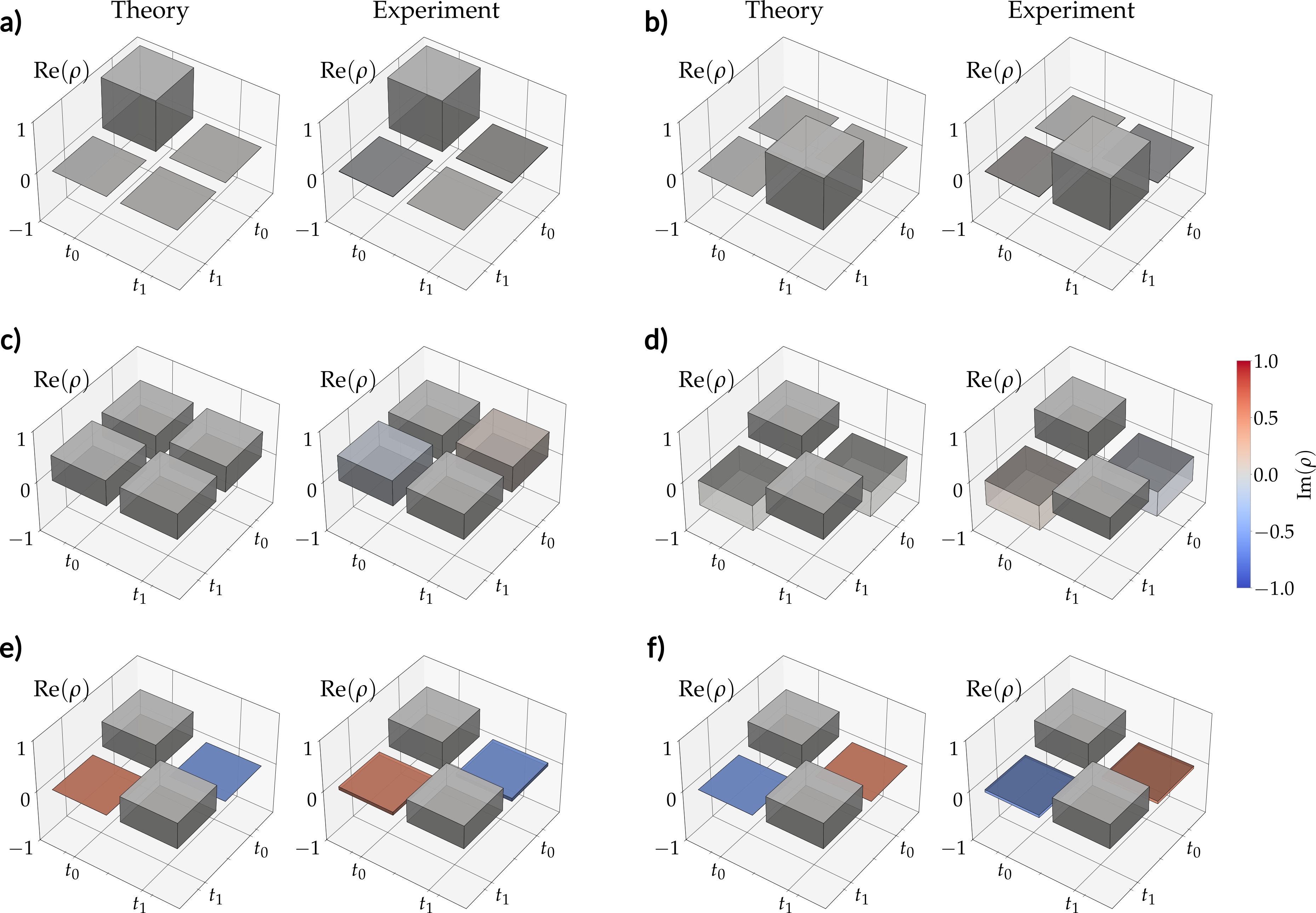}
\caption{{\bf Qubit tomography results.}
Comparison between theory and experimental results for the 6 states composing the two-dimensional MUBs: \textbf{a)} $\ket{t_0}$, \textbf{b)} $\ket{t_1}$, \textbf{c)} $(\ket{t_0}+\ket{t_1})/\sqrt{2}$, \textbf{d)} $(\ket{t_0}-\ket{t_1})/\sqrt{2}$, \textbf{e)} $(\ket{t_0}+i\ket{t_1})/\sqrt{2}$, \textbf{f)} $(\ket{t_0}-i\ket{t_1})/\sqrt{2}$.
The height of each bar indicates the magnitude of the real part of the density matrix.
The color of each bar corresponds to the magnitude of its imaginary component.}
\label{fig:mubtomo}
\end{figure*}

\begin{figure*}[h!]
\centering
\includegraphics[width=\textwidth]{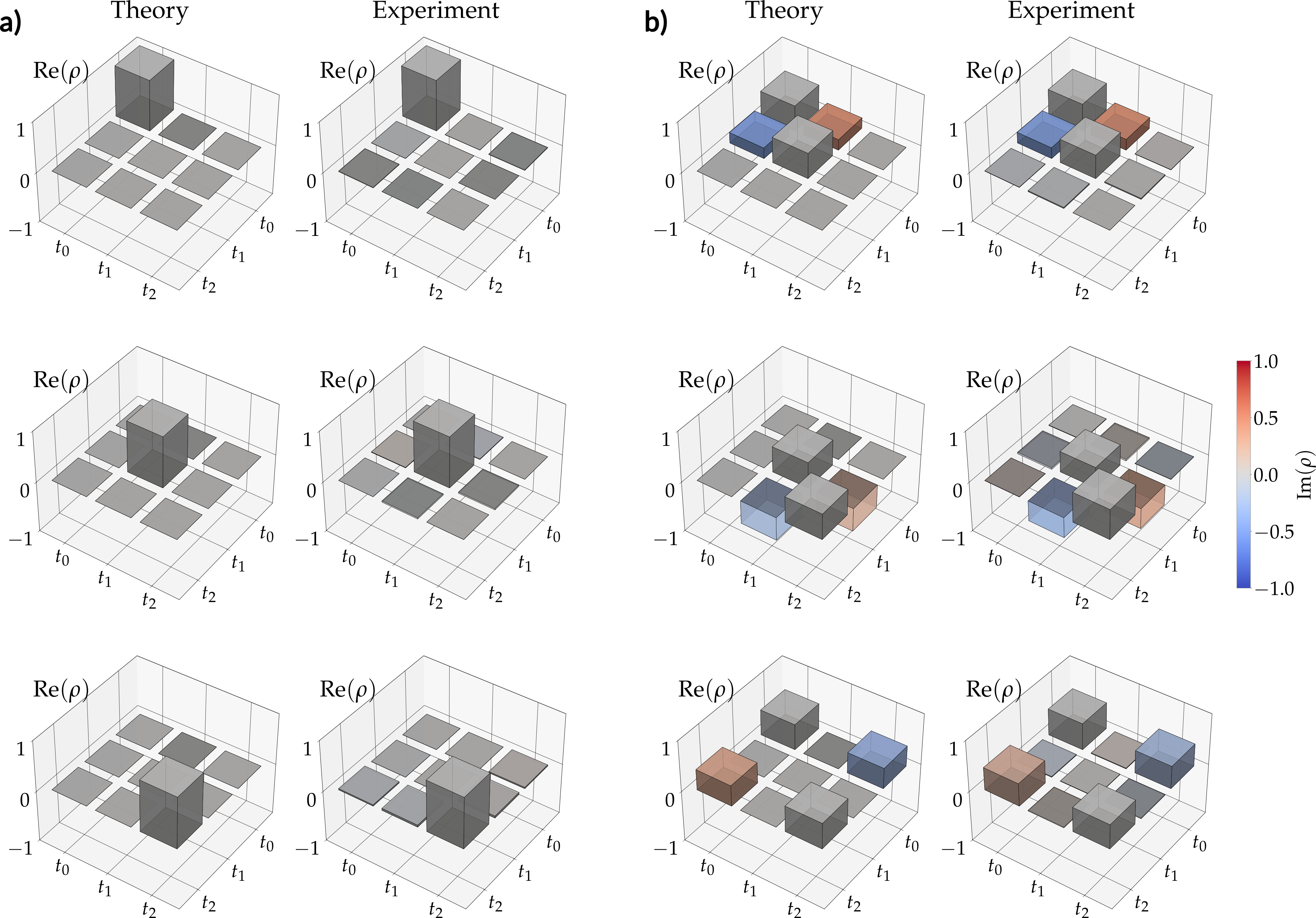}
\caption{{\bf Qutrit tomography results.}
Example comparison between theory and experimental results for various three-dimensional states: \textbf{a)} the basis states $\ket{t_0}$, $\ket{t_1}$, and $\ket{t_2}$, and \textbf{b)} three different superposition states corresponding, up to a constant relative phase, to eigenvectors of the SU(3) generators.
The height of each bar indicates the magnitude of the real part of the density matrix.
The color of each bar corresponds to the magnitude of its imaginary component.
}
\label{fig:qutrit_tomo}
\end{figure*}

We present the fidelity results for $48$ two-dimensional states (Fig.~\ref{fig:Fidelities}a), 
and for 18 three-dimensional states (Fig.~\ref{fig:Fidelities}b).
The density matrices of the unknown states are reconstructed from the experimental probabilities using maximum likelihood estimation. The fidelity of the experimental density matrix $\rho^{\rm exp}$ with respect to the theoretical state $\rho^{\rm theo}$ is defined as
$F^{\rm exp}=\text{Tr}\left[ \sqrt{\sqrt{\rho^{\rm theo}}\rho^{\rm exp}\sqrt{\rho^{\rm theo}}}\right]^2$.
In Fig.~\ref{fig:mubtomo} and Fig.~\ref{fig:qutrit_tomo}, we depict some quantum state tomography results for various experimentally prepared qubits and qutrits, respectively.

\section{V. A possible scheme for the generation and measurement of high-dimensional entangled states  using an SPDC source}
\begin{figure*}[ht!]
\centering
\includegraphics[width=0.6\textwidth]{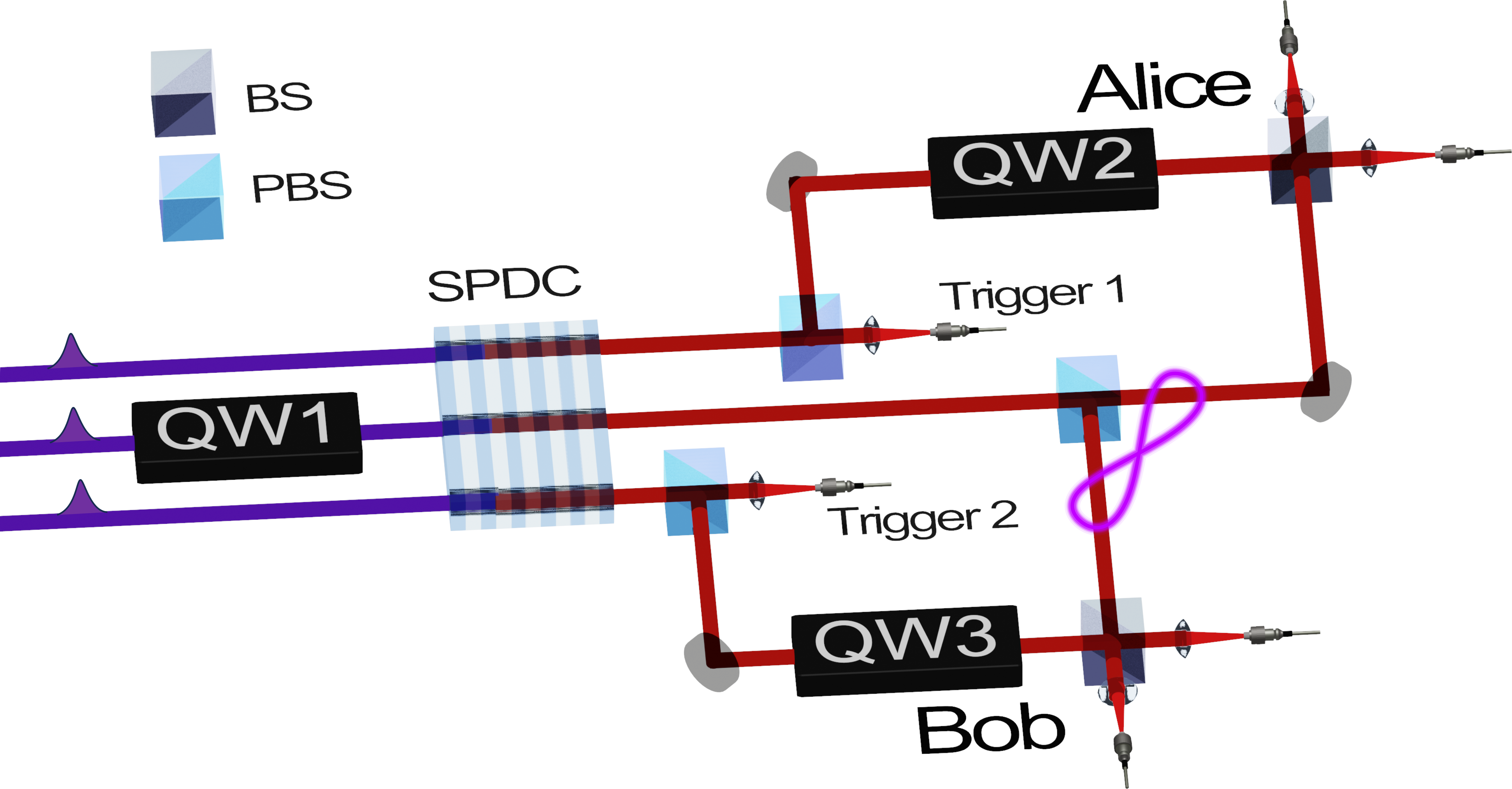}
\caption{{\bf Experimental proposal.} Scheme for generating and measuring high-dimensional entangled states using spontaneous parametric down-conversion (SPDC) generating three pairs of photons.  BS, beam splitter; PBS, polarizing beam splitter; QW, quantum walk.}
\label{fig:proposal}
\end{figure*}

The scheme theoretically proposed in the main text to generate and measure high-dimensional time-bin states can be used for entanglement-based quantum communication protocols and novel high-dimensional nonlocality tests.
We now theoretically describe a possible implementation, where photons are generated via a nonlinear crystal, as depicted in Fig.~\ref{fig:proposal}.

In this scheme, the same crystal is pumped by three parallel beams from the same laser in three different positions to produce spectrally indistinguishable single photons via type-II SPDC. Two (out of six) photons are used to herald the existence of two other reference photons that are used for measuring the entangled photons (top and bottom paths). The entangled pair (middle path) is generated by temporally shaping the pump through quantum walk dynamics.

As described in the main text, this pump laser beam can be prepared by the quantum walk dynamics (QW1 in Fig.~\ref{fig:proposal}) in the arbitrary state $ \ket{\rm pump}=\sum_i \, c_i\, \ket{t_i}$,
 with $\sum_i |c_i|^2=1$ and $\{ \ket{t_i} \}$ being the basis of time-bin space. After the nonlinear crystal, ignoring the higher-order terms, the state of the pair will be entangled and have the following form:
$\ket{\Psi}=\sum_{i=1}^n \; a_i \ket{t_i}_s\ket{t_i}_i$.

Once the entangled time-bin encoded photons are split according to their polarizations, they can be distributed among distant measurement stations, Alice and Bob. To measure the entangled pair with the HOM-measurement scheme, the heralded reference photons are prepared by the quantum walks QW2 and QW3, respectively.

It is worth mentioning that if deterministic sources reach a sufficiently high quantum efficiency, this method could be further simplified. In this scenario, Alice and Bob would generate their own reference photons synchronized with the laser pulse utilized for the entangled source at each measurement station.
The probabilistic nature of the QW protocol does not represent a fundamental limitation. One could circumvent this by preparing the reference measurement photons in a heralded manner. For example, one could herald the preparation of the state by measuring the vacuum in the second discarded output port of the QW following the final coin projection.

\clearpage

\end{document}